\documentclass[twocolumn,superscriptaddress]{revtex4-1}
\usepackage{setspace}

\usepackage{amsmath}
\usepackage{amssymb}
\usepackage{graphicx}
\usepackage{color}

\usepackage{array}
\newcolumntype{M}[1]{>{\centering\arraybackslash}m{#1}}
\newcolumntype{N}{@{}m{0pt}@{}}

\DeclareMathAlphabet\mathbfcal{OMS}{cmsy}{b}{n} 

\begin{document}

\title{Crossover behavior of the thermal conductance and Kramers' transition rate theory}

\author{Kirill A. Velizhanin}

\email{kirill@lanl.gov}

\affiliation{Theoretical Division, Los Alamos National Laboratory, Los Alamos,
NM 87545, USA}

\author{Subin Sahu}

\affiliation{Center for Nanoscale Science and Technology, National Institute of Standards and Technology, Gaithersburg, MD 20899,
USA}
\affiliation{Maryland Nanocenter, University of Maryland, Collage Park, MD 20742, USA}
\affiliation{Department of Physics, Oregon State University, Corvallis, OR 97331,
USA}

\author{Chih-Chun Chien}

\affiliation{School of Natural Sciences, University of California, Merced, CA 95343, USA}

\author{Yonatan Dubi}

\affiliation{Department of Chemistry and the Ilse Katz Institute for Nanoscale
Science and Technology, Ben-Gurion University of the Negev, Beer-Sheva
84105, Israel}

\author{Michael Zwolak}

\email{mpz@nist.gov}

\affiliation{Center for Nanoscale Science and Technology, National Institute of Standards and Technology, Gaithersburg, MD 20899,
USA}
\affiliation{Department of Physics, Oregon State University, Corvallis, OR 97331,
USA}

\begin{abstract}
Kramers' theory frames chemical reaction rates in solution as reactants overcoming a barrier in the presence of friction and noise. For weak coupling to the solution, the reaction rate is limited by the rate at which the solution can restore equilibrium after a subset of reactants have surmounted the barrier to become products. For strong coupling, there are always sufficiently energetic reactants. However, the solution returns many of the intermediate states back to the reactants before the product fully forms. Here, we demonstrate that the thermal conductance displays an analogous physical response to the friction and noise that drive the heat current through a material or structure. A crossover behavior emerges where the thermal reservoirs dominate the conductance at the extremes and only in the intermediate region are the intrinsic properties of the lattice manifest. Not only does this shed new light on Kramers' classic turnover problem, this result is significant for the design of devices for thermal management and other applications, as well as the proper simulation of transport at the nanoscale. 
\end{abstract}

\maketitle

Thermal transport is an important process in micro- and nano-scale technologies. It is often in a precarious position: On the one hand, thermal management strategies, including the engineering of low-resistance interfaces, become increasingly important as elements in electronic devices approach the atomic level. On the other hand, {\em phononics} -- phonon analogues of electronics -- seek tunability and inherently nonlinear behavior to make functional devices \cite{Li2012}. Thermal transport is thus at the forefront of nanotechnology research. Its impact in a broad array of applications has sparked advanced methods of the fabrication, control, and measurement of transport in, e.g., carbon nanotubes and single-molecule junctions~\cite{Dubi11-1,Yang_AIPA,Li2012}.

\begin{figure}[h!]
\includegraphics[width=1\columnwidth]{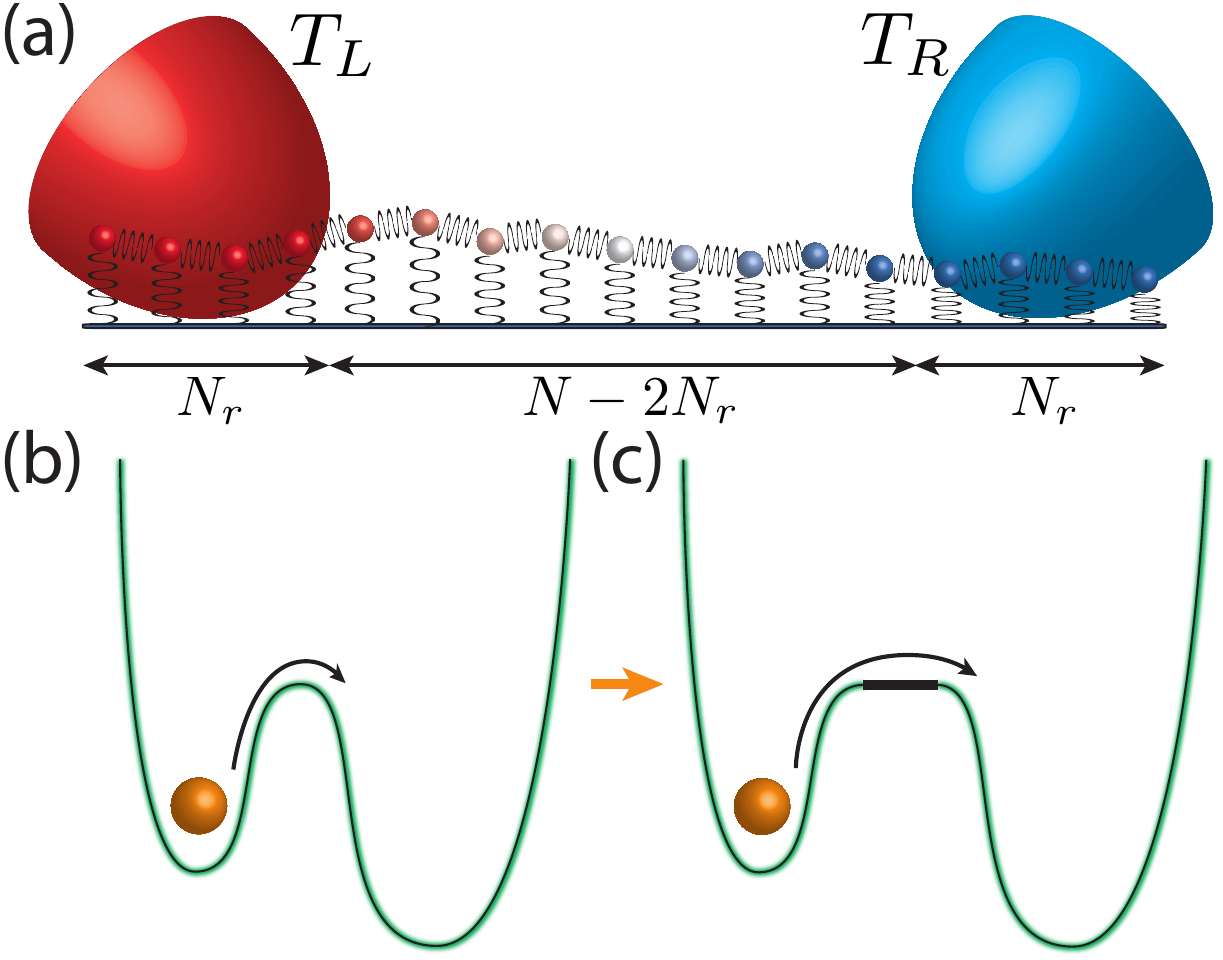} \caption{\label{fig:schematic} Thermal conductance and Kramers' transition rate problem. (a) Schematic representation of a one-dimensional 
lattice of length $N$, with $N_{r}$ sites on each side connected to independent Langevin reservoirs at temperatures $T_{L}$ and $T_{R}$. (b) Kramers' problem where noise assists the escape of a classical particle from a metastable state. The green contour of the double-well potential outlines the region where the classical particle is subject to Langevin dynamics. (c) A modified double-well potential where the barrier is deformed horizontally into a region (thick black line) where only ballistic dynamics occurs, i.e., no friction or noise.}
\end{figure}

Moreover, thermal transport is at the center of one of the major unresolved puzzles in theoretical physics, the absence of a derivation of Fourier's law of heat conduction from a microscopic Hamiltonian~\cite{Bonetto2000,Lepri_review,Dubi2009,MICHEL2006,Buchanan2005,Li2014}. This is related to the seminal work of Fermi, Pasta and Ulam (FPU) \cite{Fermi1955,Berman2005,Gallavotti2008}, which demonstrated that nonlinearity does not always lead to  thermalization. The considerations of FPU also apply to the emergence of a well-defined thermal conductivity. The role of nonlinearity -- in addition to the description of the thermal reservoirs and interfacial regions -- is thus the central topic of many studies examining thermal transport (see Refs.~\cite{Dhar_AIP08,Lepri_review,Liu2012} for recent reviews).

In this work, we demonstrate that thermal transport goes through three physically distinct regimes as the coupling to the surrounding environment -- the reservoir that supplies the heat -- changes. For weak coupling, energy input from the reservoir limits the heat current through the entire system. For strong coupling, the lattice dynamics are distorted by the presence of the reservoir and this dominates the conductance. Thermal transport is determined by the intrinsic parameters of the lattice only in the intermediate regime. These three distinct regimes exist regardless of whether the system has a well-defined conductivity or not, as we will show by theoretically studying paradigmatic examples of thermal transport~\cite{Casher71-1,Lepri_review,Dhar_AIP08,Bernardin2005,Nakazawa1970}. 
The dependence of the thermal conductance on the coupling strength to the reservoirs is physically analogous to the friction-induced turnover in Kramers' reaction rate problem for chemical reactions in solution.

\section*{\bf Results}
The first model we examine is a classical one-dimensional (1D) uniform lattice of $N$ harmonic oscillators. The Hamiltonian is 
\begin{equation}
H=\frac{1}{2}\sum_{n=1}^{N}m\dot{x}_{n}^{2}+\frac{D}{2}\sum_{n=1}^N x_{n}^{2}+\frac{K}{2}\sum_{n=0}^{N}(x_{n}-x_{n+1})^{2},\label{eq:Hamiltonian}
\end{equation}
where $x_n$ is the coordinate of the $n^{\rm th}$ oscillator, $m$ is the mass, $D$ is the strength of the on-site potential, and $K$ is the nearest-neighbor coupling constant. The entire lattice is split into three regions: $N_{r}\ge 1$ sites on the left end ($L$) and on the right end ($R$) serve as {\em extended reservoirs} as each site (i.e., oscillator) in these regions is coupled to its own ``external" Langevin reservoir of temperature $T_{L}$ and $T_{R}$ , see Fig.~\ref{fig:schematic}(a). The friction coefficient $\gamma$ gives the strength of the coupling to the external reservoirs and is taken to be the same on the both ends for simplicity. The remaining $N-2N_r$ sites in the middle comprise the {\em free lattice} ($F$). This is a generalization of a widely used model that sets $N_r\equiv 1$ \cite{Casher71-1,Dhar01-1,Dhar_AIP08}. 
In addition to being a prototypical model of thermal transport, it is also relevant to realistic systems, such as the  high- and low-temperature limits of coarse-grained models of DNA \cite{Velizhanin11-1,Chien_NT13}, where the on-site potential term represents the binding of interstrand base pairs.

\begin{figure}
\includegraphics[width=1\columnwidth]{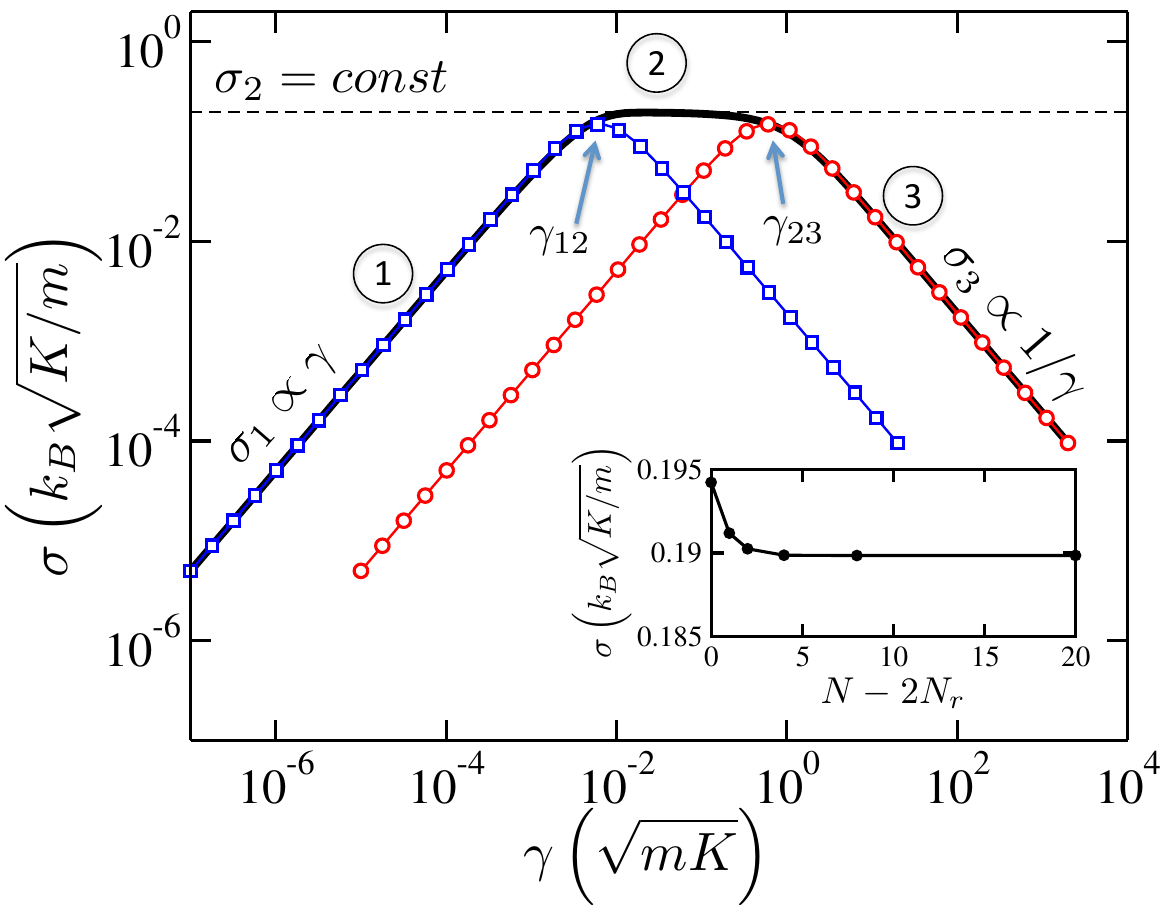} \caption{\label{fig:general_num} The thermal conductance of a harmonic lattice with $D/K=1$, $N_{r}=100$, and $N \rightarrow \infty$. The numerical results are shown as the thick black line. The analytical results for $N_r\equiv1$, Eq.~(\ref{eq:CLformula}), are plotted as $\sigma_{CL}(\gamma)$ (red circles) and $\sigma_{CL}(N_r\gamma)$, i.e., the horizontal axis is scaled with $N_r$ (blue squares). The inset shows the dependence of $\sigma$ on $N-2N_r$ at $\gamma=0.1 \, \sqrt{mK}$.}
\end{figure}

Figure~\ref{fig:general_num} shows the thermal conductance, defined as the heat current divided by $\Delta T=T_L-T_R$, for a long harmonic lattice with $N_r=100$. A detailed description of the calculation is given in the supplementary information. Three qualitatively distinct regimes, which are labeled (1), (2) and (3), are apparent in the figure. We start by examining regime (2), where the conductance depends on $\gamma$ only very weakly. The magnitude of the conductance on this plateau coincides with the {\em intrinsic conductance} of the lattice. For the anharmonic lattice we consider later on, the intrinsic conductance behaves as $\sim 1/N$ at large $N$, so there is a well-defined intrinsic {\em conductivity} in this plateau region, defined as the conductance multiplied by $N$ in the limit $N\rightarrow\infty$. 

{\bf Intrinsic conductance.}  We find the {\em intrinsic conductance} of the harmonic lattice by solving an auxiliary problem of the heat current between two semi-infinite ballistic (i.e., no friction) lattices. These lattices are initially disconnected and equilibrated at respective temperatures $T_{L}$ and $T_{R}$. When connected, the heat current flowing from left to right is $J_{L\rightarrow R}=k_B T_{L}l^{-1}_L\sum_{q>0}v_{q}$, where $v_{q}$ is the group velocity of a phonon with momentum $q$ (restricted to phonons moving to the right, i.e.,  $q>0$), $k_B$ is Boltzmann's constant, and $l_L$ is the length of the left lattice. Essentially, this expression is just the amount of classical energy stored in a specific phonon mode in equilibrium ($k_B T_L$), multiplied by its group velocity ($v_q$) and the local phonon density of states ($1/l_L$). Similarly introducing $J_{R\rightarrow L}$, taking the limit $l_L,l_R\rightarrow\infty$, and defining the total current as $J=J_{L\rightarrow R}-J_{R\rightarrow L}$, we obtain the intrinsic conductance as
\begin{equation}
\sigma_2\equiv\frac{J}{\Delta T} =\frac{k_B}{2\pi}\int_{\Omega}dq\, v_{q} =\frac{k_B}{2\pi}\int_{\Omega}dq\,\frac{d\omega(q)}{dq} =\frac{k_B\Omega}{2\pi},\label{eq:sigma_free}
\end{equation}
where $\Omega=\omega_{\rm max}-\omega_{\rm min}$ is the phonon bandwidth. This conductance is also expressible as $\sigma_{2}=\bar{v}_{q}/2$, where $\bar{v}_{q}$ is the group velocity averaged over the entire band, which gives additional insight into its form. 
The eigenmodes of the Hamiltonian, Eq.~(\ref{eq:Hamiltonian}), obey $\omega^{2}=\left[D+4K\sin^{2}(q/2)\right]/m$, 
so that $\omega_{\rm min}=\sqrt{D/m}$, $\omega_{\rm max}=\sqrt{(D+4K)/m}$, and, therefore, the phonon bandwidth is
\begin{equation}
\Omega=\sqrt{\frac{K}{m}}\left[\sqrt{D/K+4}-\sqrt{D/K}\right].
\end{equation}
Accordingly, the intrinsic conductance becomes
\begin{equation}
\sigma_2=\frac{k_B}{2\pi} \sqrt{\frac{K}{m}}\left[\sqrt{D/K+4}-\sqrt{D/K}\right].
\end{equation}
The intrinsic conductance is the maximum possible conductance of a harmonic system between two equilibrium reservoirs at different temperatures: No matter how much energy the reservoirs can pump into the system, the free lattice itself can not 
usher the energy from source ($L$) to sink ($R$) faster than the rate allowed by its intrinsic conductance. The magnitude of $\sigma_2$ is plotted in Fig.~\ref{fig:general_num} (thin dashed line), showing an excellent agreement with the numerically calculated conductance in regime (2) (hence the subscript of $\sigma_2$). We also note that decreasing $D$ -- the onsite confining potential -- and holding all other factors fixed increases the conductance, an effect that is observed in models of DNA denaturation~\cite{Velizhanin11-1,Chien_NT13}. This is purely due to the increasing bandwidth.

{\bf Casher-Lebowitz formula.} The other two regimes, (1) and (3), become physically transparent when using the Casher-Lebowitz formula \cite{Casher71-1} for a {\em single-site reservoir} at each end (i.e., $N_r\equiv 1$)
\begin{equation}
\sigma_{CL}(\gamma)=\frac{k_B\gamma}{2\pi m}\int_{0}^{2\pi}dq\,\frac{\sin^{2}q}{1+\frac{2\gamma^{2}}{mK}\left[1+\frac{D}{2K}-\cos q\right]}.\label{eq:CLformula}
\end{equation}
It is readily seen that the small and large $\gamma$ expansion of the expression yields $\sigma_{CL}\propto\gamma$ and $\sigma_{CL}\propto 1/\gamma$, respectively. In fact, very general perturbative arguments suggest that these $\sigma \propto\gamma$ and $\sigma \propto 1/\gamma$ regimes are generic for arbitrary harmonic lattices, as we show in the supplementary information. 

Equation (\ref{eq:CLformula}) is plotted in Fig.~\ref{fig:general_num} as a function of two different arguments: the bare friction coefficient, $\sigma_{CL}(\gamma)$ (red circles), and the scaled friction coefficient, $\sigma_{CL} (N_r\gamma)$ (blue squares). As is seen in the figure, they coincide with numerical results at large and small $\gamma$, respectively, which we will now explain.
 
{\bf Small $\gamma$ regime.} The small $\gamma$ expansion of Eq.~(\ref{eq:CLformula}) is $\sigma_{CL}=\frac{k_B\gamma}{2\pi m}\int_{0}^{2\pi}dq\,\sin^{2}q=\frac{k_B\gamma}{2m}$. This expression reflects the fact that energy is pumped in/out of the system via the end sites at a rate proportional to $\gamma$, and this rate is much smaller (at small $\gamma$) than the one the free lattice can carry {\em intrinsically}, $\sigma_2$. At these conditions (i.e., $k_B \gamma/m\ll\sigma_2$), the heat input by the noise term in Eq.~(\ref{eq:eom}) is so inefficient that the system has enough time to equilibrate at some global temperature $T^\prime$. 
In this quasi-equilibrium state the noise-induced heat current entering the left reservoir is $J_{L}=\frac{\gamma}{m}k_B(T_{L}-T')$, and similarly for the right reservoir $J_{R}=\frac{\gamma}{m}k_B(T_{R}-T')$. At the steady state $J=J_L=-J_R$, which yields $T'=\frac{T_{L}+T_{R}}{2}$, so that the conductance of the entire system becomes $\sigma=\frac{k_B\gamma}{2m}$, in full agreement with the small-$\gamma$ expansion of Eq.~(\ref{eq:CLformula}) above. 

\begin{figure}
\includegraphics[width=1\columnwidth]{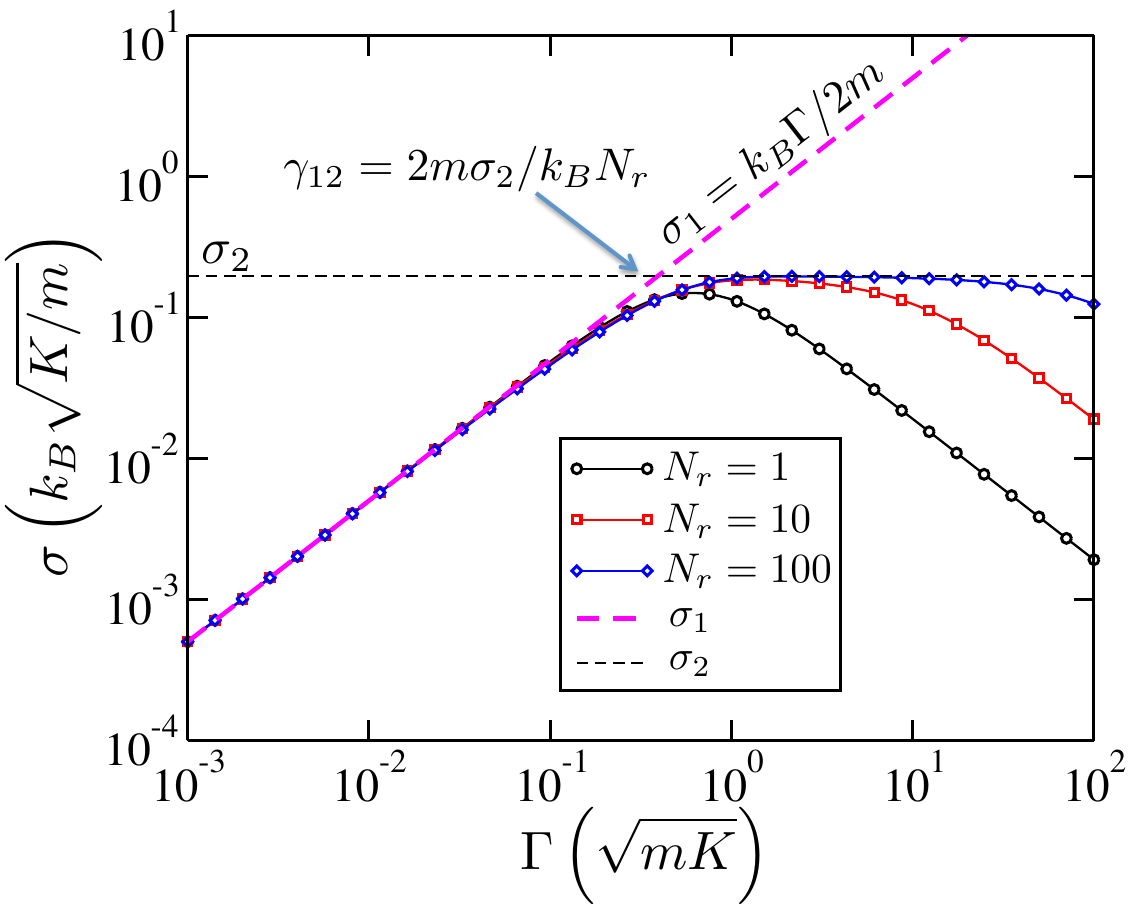}

\caption{\label{fig:weak-gamma} Comparison of numerical (circles, square, diamonds) and analytical results (dashed lines) in the small
$\Gamma$ regime at $D/K=1$ for different sizes of extended reservoirs, $N_r$.}
\end{figure}

These quasi-equilibrium-based considerations can be straightforwardly generalized to the case of extended reservoirs of arbitrary size $N_r$ yielding
\begin{equation}
\sigma_1(\gamma)=\frac{k_B N_r \gamma}{2m}\equiv \frac{k_B\Gamma}{2m},\label{eq:sigma_lowG}
\end{equation}
 i.e., the heat is pumped into the system at a rate proportional to cumulative friction constant $\Gamma=N_r\gamma$. This is the reason why Casher-Lebowitz formula (\ref{eq:CLformula}), once plotted as a function of the {\em cumulative} friction constant, $\sigma_{CL}(N_r\gamma)$, coincides with the numerical results in Fig.~\ref{fig:general_num} at small $\gamma$. The agreement between Eq.~(\ref{eq:sigma_lowG}) and numerical results in the weak friction regime is emphasized in Fig.~\ref{fig:weak-gamma} where conductances for a range of $N_r$ exactly overlaps with $k_B\Gamma / 2m$ at small friction. 
 
Eq.~(\ref{eq:sigma_lowG}) is only valid at sufficiently small $\Gamma$. When $\frac{k_B\Gamma}{2m}$ becomes higher than $\sigma_2$, the free lattice cannot transfer all the heat the Langevin reservoirs are able to supply. At that point, i.e., $\gamma \approx \gamma_{12}=2m\sigma_2/k_B N_r$ (marked in Figs.~\ref{fig:general_num} and \ref{fig:weak-gamma}), the conductance of the overall system levels off and regime (2) is established. 

{\bf Large $\gamma$ regime.} The large $\gamma$ expansion of Eq.~(\ref{eq:CLformula}) is
\begin{equation}
\sigma_3=\frac{k_B K}{2\gamma}\left(1-\frac{\sqrt{mD}}{2K}\Omega\right).\label{eq:CL_largeg}
\end{equation}
At large $\gamma$, the noise term in Eq.~(\ref{eq:eom}) can efficiently supply heat to a lattice site. However, this site is effectively decoupled from the adjacent sites since the very large friction {\em overdamps} its motion resulting in the reservoir site becoming ``off-resonant" from the rest of the lattice. We will discuss this in detail in the next section. Heuristically, however, the efficiency of heat transfer from the reservoir site to the free lattice
has to be proportional to a phonon group velocity and to a time the reservoir site can stay coherent with the rest of the lattice. The latter is given by the decoherence time of an overdamped oscillator, $1/\gamma$, so the conductance becomes (averaging over all phonon modes) $\sigma_3 \propto \frac{1}{\gamma}\int_\Omega d\omega\, v_{q}(\omega)$.
This coincides with Eq.~(\ref{eq:CL_largeg}) up to a constant prefactor. 

This heuristic argument is directly applicable to the case of extended reservoirs ($N_r > 1$). Indeed, the reservoir site directly connected to the free lattice is the one that excites phonons in the free lattice and the remaining $N_r-1$ reservoir sites (on the left or on the right) are decoupled from the lattice by at least another order of $1/\gamma$. This entails that the heat conductance is independent of $N_{r}$ in the large-$\gamma$ regime,  which agrees with the numerical results in Fig.~\ref{fig:general_num}. When $\sigma_3(\gamma)$ reaches $ \sigma_{2}$ as $\gamma$ decreases to $\gamma \approx \gamma_{23}=\frac{k_B K}{2\sigma_2}\left(1-\frac{\sqrt{mD}}{2K}\Omega\right)$ (marked in Fig.~\ref{fig:general_num}), there is a crossover from regime (3) to (2). That is, as $\gamma$ decreases the reservoir site becomes coherent enough to efficiently transfer heat to the free lattice, so the finite intrinsic conductance of the free lattice, $\sigma_2$, becomes the limiting factor.

{\bf The plateau and tilt.} Regime (2)  is just a crossover point at $N_r=1$ [Eq.~(\ref{eq:CLformula})], but it becomes a pronounced plateau at $N_r\gg 1$ as shown in Fig.~\ref{fig:weak-gamma}. This is because the position of the crossover between regimes (2) and (3), $\gamma_{23}$, does not depend on $N_r$, but the one between regimes (1) and (2), does as $\gamma_{12}\propto 1/N_r$. At sufficiently large $N_r$ the two crossovers are thus well separated, establishing a plateau in between. 

\begin{figure}
\includegraphics[width=1\columnwidth]{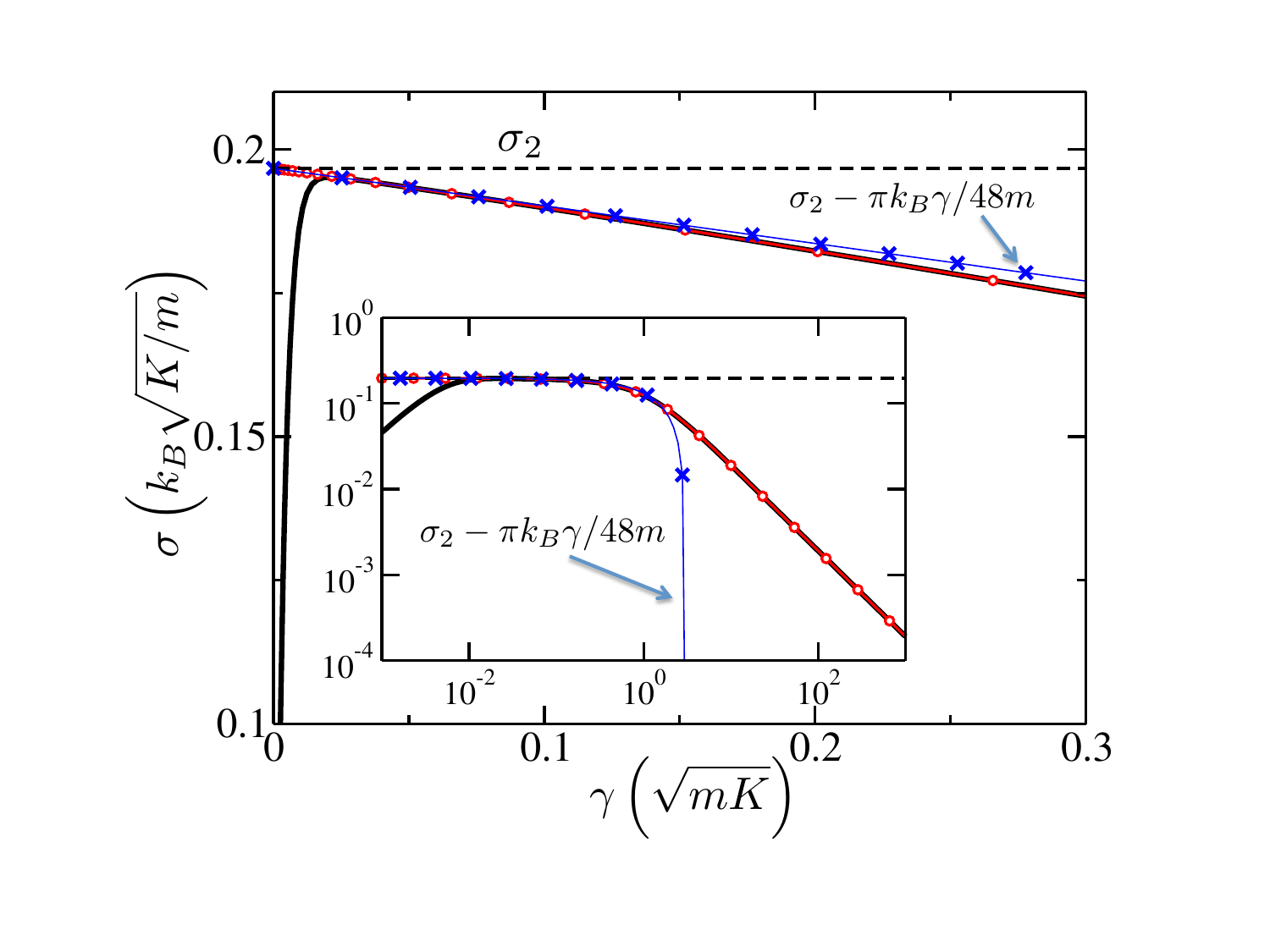}
\caption{\label{fig:reflection} Comparison of the exact numerical results (thick black line) for the conductance with the semi-analytical result,
Eq.~(\ref{eq:sigma_reflection}), shown by red circles. The model parameters are $D/K=1$ and $N_{r}=100$. Thin dashed line shows $\sigma_{2}$. The inset shows  that the semi-analytic result also works in the large $\gamma$ regime. For weak friction, Eq.~(\ref{eq:sigma_reflection}) reduces to Eq.~(\ref{eq:tilt}), shown by blue crosses.
}
\end{figure}

Closer inspection of the numerical results reveals that the ``plateau'' is in fact tilted, see Fig.~\ref{fig:reflection}. The linear fit of this tilted plateau yields an intercept with the $\gamma=0$ axis numerically close to $\sigma_{2}$ and the slope independent (within the fitting accuracy) of any parameters of the lattice.

We demonstrate below that once $N_{r} \to \infty$, the conductance becomes $\sigma_{2}$ as $\gamma \to 0$ (the order of taking the two limits is important: the limit of $N_r\rightarrow\infty$ is assumed to be taken first, i.e., $\Gamma=N_r\gamma\rightarrow\infty$). 
When $\gamma$ is finite, though, the conductance falls off linearly with $\gamma$ with a {\em universal} slope. 
This tilt can be understood via the inhomogeneity of the lattice. For example, connecting lattices that have only partially overlapping phonon bands results in a 
poor conductance due to scattering at the interfaces \cite{Terraneo2002-094302}. In our situation, the phonons in the extended reservoirs and the free lattice are {\em different} because of a finite phonon lifetime in the former (due to friction) and an infinite lifetime in the latter. Within the extended reservoir, each phonon has a spread in frequency $\delta\omega \approx \gamma$. This smearing results in imperfectly overlapping bands at the band edges, which in turn
leads to a decreasing conductance as $\gamma$ increases.

The more rigorous understanding of the tilt can be achieved by considering phonon scattering at the interface between a free lattice and a lattice with uniform friction $\gamma$ ($K$ and $D$ are the same for both). 
For a phonon of frequency $\omega$ incoming from the free lattice, the solution within the free lattice is $x_{n}(t)=e^{iqn-i\omega t}+Ae^{-iqn-i\omega t}$, where $A$ is 
the reflection amplitude. The solution in the lattice with friction is $x_{n}(t)=Be^{iq'n-i\omega t}$, where $q'$ has a non-zero
imaginary component due to the finite phonon lifetime. Imposing the boundary conditions, the reflection amplitude is 
%
\begin{equation}
A(q)=-\frac{1-e^{-i(q'-q)}}{1-e^{-i(q'+q)}},
\end{equation}
where $\cos q^\prime = \cos q - i \gamma \omega / 2 K$ and $\omega=\omega(q)$ is the phonon dispersion relation for a free lattice. The two limiting cases are $A=0$ at $\gamma\rightarrow0$ and $|A|=1$ at $\gamma\rightarrow\infty$ with $0<|A|<1$ at any finite $\gamma$. This justifies the heuristic argument above that phonons with and without friction are indeed different resulting in a finite reflection amplitude at the friction/frictionless interface. At finite $\gamma$, the reflection is most significant near the band edges (i.e., within $\delta\omega \approx \gamma$ of $\omega_{\rm min}$ or $\omega_{\rm max}$).

In equilibrium at temperature $T$, the momentum-resolved current from the free lattice to the extended reservoir is $J(q)=k_B Tv_{q}\left[1-R(q)\right]$ for $q>0$, where $R(q)=| A(q) |^2$ is the reflection coefficient. It is difficult to directly evaluate the current from the extended reservoir to the free lattice, but in equilibrium this current has to fully compensate the one entering the reservoir from the free lattice. Using this correspondence, we can now write down the current from a reservoir to the free lattice even at non-equilibrium conditions. This results in the following set of balance equations for the heat current through a free lattice between two extended reservoirs,
\begin{gather}
J_{L}(q)-R(q)J(-q)=J(q),\nonumber \\
J_{R}(-q)-R(q)J(q)=J(-q),
\end{gather}
where $J(q)$ is the current within the free lattice and $J_{L(R)}$ is the current from the left (right) reservoirs. For example, the first equation states that the
current from left to right within the free lattice equals the current from the left reservoir plus the portion of ``left" current, $J(-q)$, reflected by the interface with the right reservoir. 
Solving these equations and integrating over all phonon modes, $\int_{0}^{\pi}dq\,\left[J(q)+J(-q)\right]$, 
we obtain
\begin{equation}
\sigma(\gamma)=\frac{k_B}{2\pi}\int_{0}^{\pi}dq\, v_{q}\frac{1-R(q)}{1+R(q)},\label{eq:sigma_reflection}
\end{equation}
which is similar to Eq.~(\ref{eq:sigma_free}), except for
the quotient within the integrand. This quotient is strictly positive and always less than $1$ -- it represents the non-vanishing thermal resistance of the interfaces due to phonon scattering and is known as the Kapitza resistance \cite{Kapitza41}. The small-$\gamma$ expansion of Eq.~(\ref{eq:sigma_reflection}) produces a very simple result
\begin{equation}
\sigma(\gamma) \approx \sigma_2-\frac{\pi}{48}\frac{k_B\gamma}{m},\label{eq:tilt}
\end{equation}
shown in Fig.~\ref{fig:reflection} by blue crosses.  As is seen, the universal tilt of $-\pi k_B/48m$ is independent of lattice parameters (except for mass) and in excellent agreement with numerical results in the plateau region. Furthermore, it turns out that Eq.~(\ref{eq:sigma_reflection}) evaluated at arbitrary $\gamma$ (red circles) does not only exactly reproduce the tilted plateau but also the high-$\gamma$ regime (3). In particular, the large-$\gamma$ expansion of Eq.~(\ref{eq:sigma_reflection}) results in Eq.~(\ref{eq:CL_largeg}). Therefore, Eq.~(\ref{eq:sigma_reflection}) is {\em exact} in the limit of $N_r\rightarrow\infty$ and is valid for the ``moderate to strong friction" regime. The only regime it cannot reproduce is regime (1) since the limit $N_r\rightarrow\infty$ has been taken in the scattering calculation, which stretches the plateau all the way to $\gamma=0$.

{\bf Kramers' theory.} The crossover behavior discussed above is the thermal transport analog of the Kramers' theory for the classical transition rate of a particle out of a metastable well in the presence of friction and noise \cite{Hanggi1990-251,Melnikov1991-1}, as depicted in Fig.~\ref{fig:schematic}(b) for a double-well potential. The rate constant for the transition is 
$ k(E_{b},\gamma)=k_{0}(\gamma)e^{-E_{b}/k_B T}$, where $E_{b}$ is the height of the energy barrier and the prefactor $k_{0}(\gamma)$ depends on $\gamma$. Kramers demonstrated that $k_{0}(\gamma)\sim \gamma$ when $\gamma$ is small. In general, the energy supplied by the noise to a particle in a certain well is lost either to friction or to the particle leaving this well by overcoming the barrier. The latter energy loss channel dominates when $\gamma$ is very small so the friction can be essentially neglected. At these conditions, the transition rate is limited by the low rate at which noise can supply energy needed for a particle to overcome the barrier. This rate is linear with respect to $\gamma$, resulting in $k_0(\gamma) \sim \gamma$. This behavior is indeed analogous to regime (1) in the thermal transport problem, where the free lattice is very efficient in carrying heat, so the heat current is limited by the rate at which the Langevin reservoir can input energy into the system, Eq.~(\ref{eq:sigma_lowG}). 

On the other extreme, as $\gamma$ becomes very large, Kramers showed that $k_{0}(\gamma) \sim 1/\gamma$. In this regime, the dynamics of the particle in the Kramers' problem becomes increasingly non-ballistic due to the strong friction and noise. This results in the high probability of {\em re-crossings}, i.e., even if the particle overcomes the barrier and crosses the surface separating the wells, the very strong noise can still push it back thus preventing its thermalization in the new potential well \cite{Melnikov1991-1}. The probability of re-crossings grows with $\gamma$ resulting in $k_0(\gamma) \sim 1/\gamma$.  

In order to illustrate the re-crossing phenomenon in the Kramers' problem and emphasize its similarity to the large-$\gamma$ regime (3) of the thermal transport problem, we deform the barrier in the Kramers' problem as shown in Fig.~\ref{fig:schematic}(c). Specifically, we ``stretch" the very top of the barrier into a horizontal {\em ballistic} region (i.e., no friction/noise) of a finite length (this is in contrast with the example of Section VII.E of Ref.~\cite{Hanggi1990-251}, where dissipation is present everywhere). This modification does not affect the transition rate since if a particle enters this region with a certain velocity, it will leave this region with the same velocity (remember that, within our model, there is no friction in the horizontal region). The only thing that changes by adding this ``stretching" is the time required for a particle to cross the barrier -- irrelevant in the steady state. 

Once a particle reaches the top of the barrier going from the left, it propagates freely along the ballistic region until it reaches the onset of the right well. Upon hitting this onset, the particle immediately becomes subject to noise which, if strong enough, can kick it back to the ballistic region, so the particle might end up in the the well it originally came from. Thus, the particle can be thought of as being {\em reflected} off the boundary between ballistic and non-ballistic regions. This is the phenomenon of re-crossing -- the top of the barrier can be crossed multiple times without thermalization in either of the potential wells.    

This perspective demonstrates the analogy to thermal transport at ``moderate to strong" $\gamma$. Indeed, we were able to describe the thermal transport in regime (2) and regime (3), see inset in Fig.~\ref{fig:reflection}, by considering {\em reflections} of phonons off the boundary between the free lattice (no friction) and one of the reservoirs (friction is present). The ballistic region and the two potential wells in the deformed double-well potential are then respective analogs of the free lattice and the extended reservoirs in the thermal transport problem. 

This deep physical similarity between the two problems, when looked upon from the perspective of particle (or phonon) reflection, calls for qualitatively similar behavior as the magnitude of friction varies. Indeed, Eq.~(\ref{eq:sigma_reflection}) gives $\sigma \sim 1/\gamma$ at large $\gamma$, which matches $k_0(\gamma) \sim 1/\gamma$ in Kramers' problem. Furthermore, at  intermediate values of $\gamma$, the general Kramers' solution reduces to the transition state theory (TST) rate $k_{\rm TST}$ which does not depend on $\gamma$ \cite{Hanggi1990-251,Melnikov1991-1}. More accurately, the Kramers' rate equals to $k_{\rm TST}$ with negative corrections linear with respect to $\gamma$, so that $k_{\rm TST}$ is always an exact upper limit of the Kramers' rate. This happens as well in the thermal transport problem where the conductance in regime (2) is given by Eq.~(\ref{eq:tilt}). In this equation, $\sigma_2$ -- an exact upper limit  -- is an {\em intrinsic} conductance of the free lattice, which does not depend on friction. 

We note that similar arguments can be applied to a version of the Kramers' problem where a classical particle escapes a single metastable potential well \cite{Hanggi1990-251}. In this formulation, Kramers' problem becomes analogous to a problem of thermal transport through the interface between a free lattice and a lattice with uniform friction studied by us when discussing the tilt of the plateau. In particular, the low probability of phonon transmission through the interface between these two lattices at large $\gamma$ is analogous to the particle escape rate scaling as $1/\gamma$ in the Kramers' problem of a metastable potential well. 

The appealing picture developed above is based on very general and intuitive physical arguments and is, therefore, expected to be valid beyond the specific case of a uniform harmonic lattice. Indeed, below we discuss the two important cases of (i) a harmonic lattice with disorder and (ii) an anharmonic lattice. We demonstrate the existence of three distinct regimes of thermal transport and, therefore, the similarity to Kramers' problem. 

{\bf Disordered Harmonic Lattice.}\label{sec:harm_disorder}
The Hamiltonian for a harmonic lattice with mass disorder reads as
\begin{equation}
H=\frac{1}{2}\sum_{n=1}^{N}m_n\dot{x}_{n}^{2}+\frac{D}{2}\sum_{n=1}^N x_{n}^{2}+\frac{K}{2}\sum_{n=0}^{N}(x_{n}-x_{n+1})^{2}.\label{eq:HamDis}
\end{equation}
The mass disorder is realized by sampling mass uniformly and independently for each lattice site within the interval $m_n=m\pm\delta m=m(1.0\pm0.3)$, so that the mean mass, $m$, is the same as in the non-disordered case, Eq~(\ref{eq:Hamiltonian}). The central ``free" part of the lattice is kept to be $N-2N_r=60$ sites long. The sizes of the extended reservoirs are $N_r=$1, 10 and 100 (the same on the left and on the right in each calculation).

\begin{figure}
\includegraphics[width=1\columnwidth]{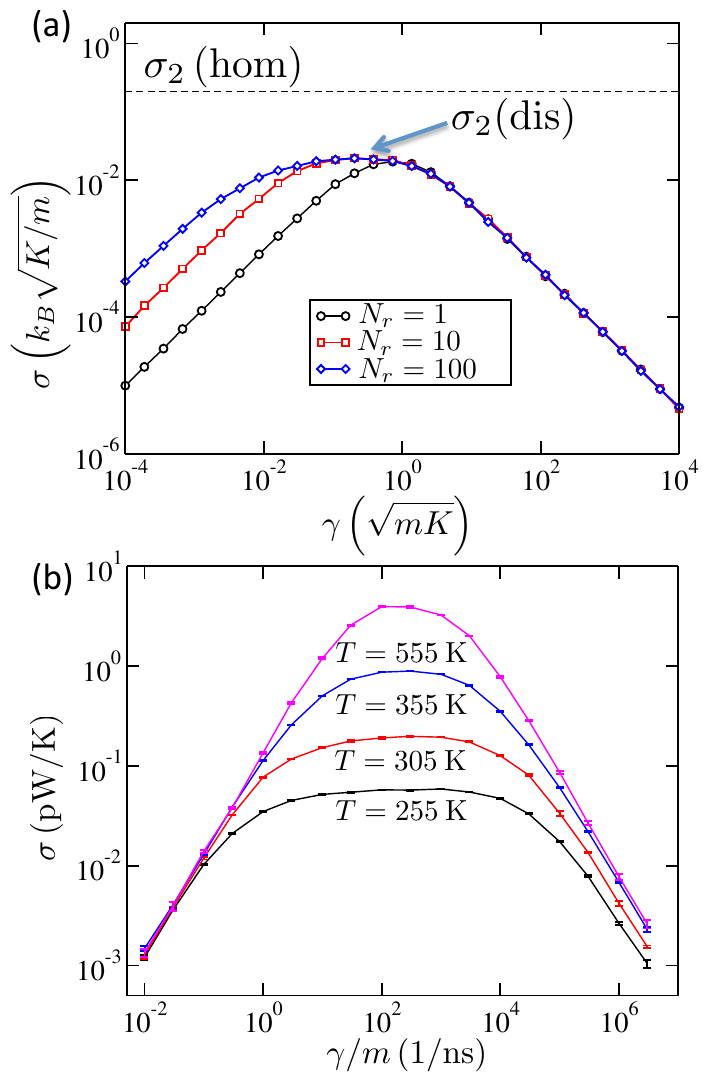}
\caption{\label{fig:disorder_PBD} Crossover behavior in disordered and anharmonic lattices. (a) The thermal conductance of a disordered harmonic lattice with $N_r=$1, 10 and 100 (black circles, red squares and blue diamonds) and $N-2N_r=60$. The system is the same as that in Fig.~\ref{fig:general_num}, i.e., $D/K=1$, except for the mass which is now sampled uniformly and independently for each site within the interval $m_n=m(1\pm0.3)$. The dashed line gives the intrinsic conductance for a homogeneous lattice, $\sigma_2({\rm hom})$, demonstrating that the disorder significantly decreases the heat conductance of the lattice, $\sigma_2({\rm dis})$ at the plateau. The curve is averaged over $100$ to $1000$ realizations of the disorder so that magnitude of the error due to the incomplete averaging over disorder is comparable to the thickness of the lines.
(b) The thermal conductance of anharmonic PBD lattice with $N_r= 20$ and $N = 100$ for various average temperatures $T$. The error bars are computed using the range of fluctuations of the cumulative current for the final 10~\% of the simulation time.}

\end{figure}

Figure~\ref{fig:disorder_PBD}(a) shows the heat conductance versus $\gamma$ for a harmonic lattice with mass disorder. The presented numerical results are statistically averaged over the disorder. The crossover behavior is indeed present. Furthermore, when $N_r$ increases the plateau starts to form as it was the case for the homogeneous lattice. However, the magnitude of the conductance at the plateau, $\sigma_2({\rm dis})$, is approximately an order of magnitude lower than $\sigma_2({\rm hom})$ (dashed black line), the latter being the intrinsic conductance of the homogeneous lattice. This is due to disorder-induced finite phonon mean free path being shorter than $N-2N_r=60$, which lowers the conductance. 

{\bf Anharmonic Lattice.}  To investigate the crossover behavior in an anharmonic lattice we examine the thermal conductance in a paradigmatic nonlinear lattice -- the Peyrard-Bishop-Dauxois (PBD) model  \cite{PB_PRL89,PBD_RAPID93,PBD_PRE93}. The PBD model is a one dimensional lattice that represents nonlinear fluctuations of DNA as it denatures. Here, we are primarily interested in the crossover behavior for a highly nonlinear lattice, rather than the physics of the denaturation transition where the double helix separates into two single strands (the transport physics described by this model has been discussed elsewhere, see Refs.~\cite{Velizhanin11-1,Chien_NT13}). The Hamiltonian takes on the form
\begin{equation}
H=\frac{1}{2}\sum_{n=1}^N m\dot{x}_n^2+\sum_{n=1}^N V(x_n)+\sum_{n=0}^N W(x_n,x_{n+1}),\label{eq:PBD}
\end{equation}
where $x_n$ represents the base-to-base distance within the $n^{th}$ base pair. The on-site Morse potential $V(x_n)= \tilde{D}(e^{-a x_n}-1)^2$ represents the hydrogen bonding between the bases and the coupling term $W(x_n,x_{n+1})=\frac{\tilde{K}}{2}(1+\rho e^{-b(x_n+x_{n+1})})(x_n-x_{n+1})^2$ gives the stacking interaction. We note that with small amplitude fluctuations (at low temperature) this model is harmonic with parameters $D=2 \tilde{D}/a^2$ and $K = \tilde{K}(1+\rho)$. Similarly, at high temperature, the model is harmonic with parameters $D=0$ and $K = \tilde{K}$. In between, the model displays highly nonlinear behavior. 

The results of simulations, shown in Fig.~\ref{fig:disorder_PBD}(b), clearly demonstrate three regimes of thermal transport. Specifically, $\sigma(\gamma)\sim\gamma$ and $\sigma(\gamma)\sim 1/\gamma$ at small and large $\gamma$, respectively. Regime (2), where $\sigma(\gamma)\approx {\rm constant}$, is most pronounced at lower temperatures because the extended reservoir size is kept fixed and the conductance is lowest at this point, see the discussion below. We have shown elsewhere that this regime has a well-defined conductivity~\cite{Chien_NT13}.

Unlike the harmonic model, the intrinsic conductance of nonlinear lattices does in general depend on temperature. In case of the PBD model, the intrinsic conductance of the lattice increases as the average temperature increases across the denaturation transition, which is apparent in the increase of the conductance plateau. This increase is due to the decrease in nonlinearity of the model as the transition is crossed from below. Moreover, the length of the plateau region shortens due to the increasing conductance. At higher temperatures, the crossover from Regime (1) to Regime (2) happens at a larger $\gamma$, as the rate of heat input from the reservoir ($\sim \gamma$) has to compete with the intrinsic conductance of the complete reservoir/lattice system. A similar shift in the crossover happens for Regime (2) to Regime (3).  In addition to this model, the crossover behavior has been observed in other anharmonic lattices, such as the FPU lattice~\cite{Lepri_review}. We note that for anharmonic lattices the heat conductance can scale nontrivially with the length and temperature, which will result in an intricate interplay between these variables, the extended reservoir size, and the width of the plateau region, as seen in Fig.~\ref{fig:disorder_PBD}(b).

\section*{Discussion}
We elucidated the mechanisms behind the crossover behavior of thermal transport  as the strength of coupling to the reservoirs is varied. The evidence suggests that this behavior is universal, 
applying to harmonic, anharmonic, and disordered systems. It is also guaranteed to exist in higher dimensional harmonic systems as well, due to the closed form expression of transport in arbitrary 
lattices. This phenomenon parallels the physical behavior observed in Kramers' transition state problem. Our results illuminate the regime where the intrinsic conductance is manifest. It is in this regime where nonlinear fluctuations, disorder, etc., dominate the conductance and where thermal transport can be used to probe physical processes, such as DNA denaturation~\cite{Chien_NT13,Velizhanin11-1}.  We also note that in many physical systems, the friction coefficient $\gamma$ depends on frequency, giving rise to memory in the equations of motion \cite{Banerjee2006,Dhar_AIP08,Dhar01-1}. We expect a crossover to still occur when the overall coupling to the external reservoirs is tuned. The intermediate regime, however, may display more complex behavior due to how the reservoirs affect modes at different frequency scales. We leave this study for a future investigation.

Moreover, non-equilibrium molecular dynamic simulations is the standard tool in the study of thermal transport in nanoscale systems (see, 
e.g., refs \cite{Luo2010,Wang2012,Zhang2010,Falat2011,Yang_AIPA,Wang2009,Saha2007,Xu2014,Mortazavi2012}). In these simulations the strength of coupling to the environment (in the form of, e.g., Langevin or Nose-Hoover 
thermostats) is a free parameter. It is relatively innocuous when a well-defined conductivity exists. However, when both ballistic and diffusive effects are present -- as is the case at the nanoscale -- the choice of this coupling affects the calculation of the thermal conductance. It thus must be chosen to appropriately simulate the property of interest, whether it is the thermal conductance of the device or the intrinsic conductance of the functional system. This is especially important when extracting scaling exponents of the conductance versus temperature or lattice length in nonlinear and disordered systems, as both quantities nontrivially affect the conductance and the crossover behavior can spuriously influence the predicted scaling.

\section*{Methods}
{\bf Harmonic Lattices.} The harmonic lattices we consider are described by Eq.~(\ref{eq:Hamiltonian}). We also connect two additional sites $x_0$ and $x_{N+1}$ to the ends of the lattice, which are fixed at zero. The Langevin equations of motion are
\begin{equation}
m\ddot{x}_{n}+Dx_{n}+K[2x_{n}-x_{n-1}-x_{n+1}]=f_{n}(t),\label{eq:eom}
\end{equation}
where the l.h.s.\ describes the Hamiltonian dynamics and the r.h.s.\ is the reservoir-induced noise and friction forces, $f_{n}(t)=\eta_{n}(t)-\gamma_{n}\dot{x}_{n}(t)$. The noise $\eta_n (t)$ and friction coefficient $\gamma_{n}$ are related by the fluctuation-dissipation relation
\begin{equation} 
\langle\eta_{n}(t)\eta_{m}(t')\rangle=2\gamma_{n}k_B T_{n}\delta_{nm}\delta(t-t').
\label{eq:force_corr}
\end{equation} 
The three regions of the lattice ($L$, $F$ and $R$) are encoded in equations of motion by setting
\begin{equation}
\gamma_n=\left\{ \begin{array}{l}
\gamma,~~ n \in L \, \textrm{or} \, R \\
0, ~~n\in F \end{array} \right.
\end{equation} 
and $T_{n}=T_{L}$ or $T_R$ for $n \in L$ or $R$, respectively. Here, a collection of independent (uncorrelated) single site Langevin reservoirs approximates the contact to the thermal reservoir \cite{segal2003}.
When $T_{L}=T_{R}=T$, the lattice will relax into an equilibrium state at temperature $T$,
as guaranteed by the fluctuation-dissipation theorem. When $T_{L}\neq T_{R}$, a heat current will flow (we assume $T_{L}>T_{R}$ without loss of generality).

The current flowing from site $n$ to $n+1$ is given by \cite{Casher71-1,segal2003}
\begin{equation}
J_{n}=K\langle\dot{x}_{n+1}x_{n}\rangle,\label{eq:current}
\end{equation}
where the average is over the statistical ensemble. For $n \in F$,  $J_{n}$ will be independent of $n$, i.e., $J_{n}\equiv J$, in the steady state. 

We note that thermal transport occurs ballistically in a frictionless harmonic lattice. This results in a diverging {\em conductivity}, which can be defined as $\lim_{N\rightarrow\infty}(N-2N_r)\sigma$ at fixed $N_r$. However, the conductance, $\sigma$, is well defined and is rapidly converging to its value corresponding to an infinite lattice, as shown in the inset of Fig.~\ref{fig:general_num}. Henceforth, we always assume $N\rightarrow\infty$ at fixed $N_r$ for homogeneous harmonic lattices. Dimensional analysis shows that the conductance of a uniform harmonic lattice in the limit $N\rightarrow\infty$ takes the form (see the Supplementary Information for the derivation)
\begin{equation}
\sigma=\frac{J}{\Delta T}=k_B\sqrt{\frac{K}{m}}C\left(\frac{D}{K},\frac{\gamma}{\sqrt{mK}},N_r\right),
\end{equation}
where $\Delta T = T_L - T_R$ and $C$ is a as of yet unknown dimensionless function of three dimensionless arguments. This expression implies that harmonic lattices with distinct values of parameters are nevertheless physically similar if $\sqrt{K/m}$, $D/K$, $\gamma/\sqrt{mK}$, and $N_r$ are identical. Such models can be said to form a ``similarity class". In what follows, all the figures pertaining to the harmonic case are plotted in natural units, i.e., $\gamma$ is plotted in units of $\sqrt{mK}$, $\sigma$ is plotted in units of $k_B\sqrt{K/m}$, etc. Note that we use $\Delta T$ defined with the external reservoir temperatures $T_L$ and $T_R$, as these are the ones typically set/measured experimentally.

For the anharmonic lattice, the setup is the same as for harmonic lattices, but with the equation of motion given by its respective Hamiltonian. 

{\bf Numerical Methods.} For anharmonic lattice the evolution of the coordinates is computed using the Br\"unger-Brooks-Karplus integrator with a time step of 10 fs. The total length of the simulation varies from 0.1 ms to 2 ms depending on the convergence of current. The temperature difference between the hot and cold reservoirs was maintained at 9.3K for all calculations. The parameters used are same as in Ref. \cite{PBD_RAPID93}, i.e., $\tilde{D} = 0.04$ eV, $a = 44.5$ nm$^{-1}$, $\tilde{K} = 4$  eV/nm$^{2}$, $\rho = 0.5$, $b = 3.5$ nm$^{-1}$, and $m = 300$ u.


\section*{\bf Acknowledgements}
K.A.V. was supported by the U.S. Department of Energy through the LANL/LDRD Program. Y.D. acknowledges support from the Israel Science Fund (grant No. 1256/14). 
S. Sahu acknowledges support under the Cooperative Research Agreement between the University of Maryland and the National Institute of Standards and Technology Center for Nanoscale Science and Technology, Award 70NANB10H193, through the University of Maryland.

\section*{\bf Contributions}
M.Z. proposed the project and C.C.C. and M.Z. obtained an initial description of the crossover behavior. K.A.V. suggested the connection to Kramers’ problem and developed an elegant theory of the crossover. K.A.V., C.C.C., and M.Z. performed analytical calculations and S.S. and K.A.V. performed numerical calculations. All authors wrote the manuscript and clarified the ideas.

\section*{\bf Competing financial interests}

The authors declare no competing financial interests.

\end{document}